\documentclass[runningheads]{llncs}
\usepackage[T1]{fontenc}
\usepackage{graphicx}
\usepackage{graphicx}
\usepackage{url}
\usepackage{cite}
\usepackage{enumitem} %custom lists
\usepackage{caption} %subfigure
\usepackage{subcaption} %subfigure
\usepackage{lscape} %vertical full page tables
\usepackage{booktabs} %look of table
\usepackage{colortbl}
\usepackage{pdfpages}  % to include PDF in appendix
\usepackage{soul} %for highlighting text
\usepackage{multirow} %for spanning columns
\usepackage[table]{xcolor} %to colour cells in tables
\begin{document}
\title{Exploring User Perspectives on Data Collection, Data Sharing Preferences, and Privacy Concerns with Remote Healthcare Technology}
\titlerunning{User Perspectives on Remote Healthcare Technology}
%
%\titlerunning{Abbreviated paper title}
% If the paper title is too long for the running head, you can set
% an abbreviated paper title here
%
\author{Daniela Napoli\inst{1}
\and
Heather Molyneaux\inst{2}
\and
Helene Fournier\inst{2}
\and
Sonia Chiasson\inst{1}
}
\authorrunning{D. Napoli et al.}
% First names are abbreviated in the running head.
% If there are more than two authors, 'et al.' is used.
%
\institute{Carleton University, Ontario, Canada\\ \email{daniela.napoli@carleton.ca,} \email{chiasson@scs.carleton.ca} \\ \and
National Research Council Canada, New Brunswick, Canada\\ \email{\{heather.molyneaux,helene.fournier\}@nrc-cnrc.gc.ca}
}

\maketitle              % typeset the header of the contribution
\begin{abstract}

Remote healthcare technology can help tackle societal issues by improving access to quality healthcare services and enhancing diagnoses through in-place monitoring. These services can be implemented through a combination of mobile devices, applications, wearable sensors, and other smart technology. It is paramount to handle sensitive data that is collected in ways that meet users' privacy expectations. We surveyed 384 people in Canada aged 20 to 93 years old to explore participants' comfort with data collection, sharing preferences, and potential privacy concerns related to remote healthcare technology. We explore these topics within the context of various healthcare scenarios including health emergencies and managing chronic health conditions. %Based on our findings, we discuss contextual nuances that may impact users' needs for remote healthcare technology.

\end{abstract}
\section{Introduction}

In this paper, we define remote healthcare technology as \emph{systems of devices designed for users to manage or support their health and well-being outside of typical healthcare settings}. The scope of remote healthcare technology\footnote{Also referred to as virtual care or telemedicine.} varies widely within related literature~\cite{harrington2022healthtech}. It can include sensing and monitoring technology~\cite{adams2018keppi}, assistive technology~\cite{antona2019robot}, patient portals~\cite{sakaguchi2017portal}, video games~\cite{hall2012videogames}, or disease management tools~\cite{stellefson2013disease}. 

Overall, these types of healthcare technology can have positive societal impact. With remote healthcare technology, patients can have increased control over their efforts in maintaining their health and well-being~\cite{haleem2021telemed}. Additionally, in-situ data gathered by these technologies can better inform healthcare practitioners' decision-making processes and lead to more accurate diagnoses and more appropriate treatment plans~\cite{WHO2019summary}. Remote healthcare technology can even help minimize unnecessary clinical visits and reduce strain on health systems~\cite{WHO2019evidence}. 

However, privacy measures adopted by remote healthcare devices do not always align with users' expectations~\cite{wilkowska2011perception,berridge2022control}. Devices used for remote healthcare technology may not be governed by legislation in ways that align with users' privacy expectations~\cite{wasser2016iot}. Like other digital systems, data that is collected and stored by remote healthcare technology can be compromised, disclosed to undesired parties, sold, or manipulated in ways that are misaligned with users' intentions. These issues are especially problematic given the sensitive nature of the data.

\textbf{Contributions:} Few studies have looked at remote healthcare technology within Canada.  Furthermore, previous work exploring remote healthcare technology~\cite{frik2019privacy,ray2022olderadults} has studied data collection and data sharing in a general sense, without differentiating between different data types.  Secondly, these types of studies asked participants about healthcare broadly~\cite{Nurgalieva2020,frik2019privacy,ray2022olderadults,fournier2022iot}, or consider a single scenario~\cite{Styliadis2014}. 

In this paper, we conduct a survey with 384 people in Canada regarding their perceptions of remote healthcare technology. Two questions guide our work: \textbf{\emph{ [RQ1]} What are  users' perspectives on data collected and shared by remote healthcare technology?} and \textbf{\emph{ [RQ2]} How might perspectives and considerations vary depending on contexts such as a user's age and healthcare scenario?} We asked about participants' comfort with data collection via remote healthcare technology, sharing preferences, and data privacy concerns. The literature highlights the contextual nature~\cite{frik2020model} of users' perspectives on remote healthcare technology thus, we explored contextual nuances that could impact users' perspectives. Rather than studying data collection and sharing in general, we extend previous work by exploring how users' perspectives may vary between various data types including \emph{identifiable video data}, \emph{anonymous video data}, \emph{audio data}, \emph{vital signs data}, \emph{wellness and activity data}. Secondly, rather than asking about healthcare broadly, we extend previous work by comparing responses between four assigned healthcare scenarios: (1) screening mild symptoms of an illness, (2) experiencing a health emergency, (3) rehabilitating after an operation, and (4) managing a chronic health condition. Finally, we compare responses from participants across four age groups.

\section{Background}
Currently, remote healthcare technology is used to enhance, rather than replace, existing care services~\cite{WHO2019summary}. As the technology continues to be adopted, it is important to ensure privacy measures align with users' expectations. This topic has been explored within the context of other countries such as the United States~\cite{barkhuus2012mismeasurement,duckert2022protecting,frik2020model}, Sweden~\cite{nymberg2022trends}, Denmark~\cite{svendsen2021pros}, and Australia~\cite{nohr2017nation}. Fewer studies~\cite{braund2023exploring,daroya2024gbqm} have covered Canadian contexts.

Some key characteristics shape healthcare in Canada. For one, while the federal government sets national standards for healthcare, services are administered by individual provinces and territories. This decentralized model results in varied healthcare experiences across the country~\cite{cihi2023commonwealth}. Additionally, many healthcare costs in the country are publicly financed. All citizens, permanent residents, and individuals on certain visas can access many healthcare services for free. Some individuals must also pay out of pocket or rely on private health insurance for services which may be uncovered or only partially covered, including prescription drugs, long-term care, mental health care, dental, and vision care. Therefore, socioeconomic inequalities in Canada can result in disparities in health outcomes between sub-populations~\cite{who2020canada}. 

Recently, Canadian governments pledged \$350 million to advance digital tools and approaches for providing remote healthcare to citizens~\cite{canada2022pan}. These efforts focused on longstanding issues for Canada's healthcare system, including: providing more equitable access to services, extending health data standards and interoperability, addressing human resource challenges, and alleviating strains on patient and provider engagement~\cite{cihi2023virtual}. Remote healthcare solutions, including those funded by the government, can serve as critical tools when a growing and aging population adds pressure for publicly available, specialized health and home care services~\cite{cihi2017seniorspop,census_hallman_2022}. 

Findings in related literature may not be well suited for Canadian contexts. Nissenbaum's Contextual Integrity (CI) framework~\cite{nissenbaum2004contextual} defines Privacy as an ``appropriate flow'' of personal information that may change depending on a number of factors including \emph{personal values}, \emph{societal rules} and other \emph{norms}. CI theory has been previously applied to healthcare privacy research~\cite{duckert2022protecting,barkhuus2012mismeasurement,frik2020model}, where researchers suggest a wide variety of social (e.g., courtesy, modesty, uncertainty about the audience) and personal (e.g., privacy, personal safety) reasons influencing users' privacy perspectives on the collection and sharing of health data. In this vein, unique nuances specific to Canadians could influence their needs and privacy concerns in remote healthcare technology.

\textbf{Remote Healthcare Technology:} When exploring the literature, we focus on technologies intended for patients to supplement traditional healthcare, including commercial devices, medical devices, and prototypes proposed by researchers. Harrington et al.~\cite{harrington2022healthtech} offer a literature review covering a variety of remote healthcare research prototypes and assessing several commercially available solutions to explore the potential of these technologies in supporting older adults with their health and well-being. The review shows that the scope of remote healthcare technology varied significantly between studies. 

For example, some options intended to support daily well-being can involve general consumer devices such as smartphones, tablets, and handheld entertainment systems to help users engage in programs that support their health~\cite{zeng2018gamification,hall2012videogames}. Other technologies may include wearables~\cite{mcmahon2016fitbit,cho2015wearable} and voice-controlled assistants~\cite{trajkova2020echo} to capture data for remote healthcare purposes. 

Other options leverage networked devices to collect, assess, and store data to help provide healthcare services. These networked devices~\cite{anderson2019gait,adams2018keppi,antona2019robot} may include sensors to collect biometric and non-biometric data such as cameras capturing video to measure gait~\cite{anderson2019gait} or pressure sensors in floors and beds to detect falls~\cite{siwicki2020remotedata}. Data from more complex and pervasive remote healthcare technology provides additional context to support healthcare treatments. 

Remote healthcare technology can be used to extend care beyond hospital or in-person doctor appointments and enable users to monitor and help assess their own healthcare needs from home~\cite{blount2007ibm}. Technology applied for this purpose may leverage a variety of sensors to collect biometric data that can include data like heart rhythm, pulse, respiratory rate, oxygen levels, blood pressure, or glucose levels~\cite{olmedo2022remoteelderly}.

In the examples mentioned so far, users can actively interact with remote healthcare technology to intentionally provide data that will be used to inform their healthcare services. With other forms of remote healthcare technology, users play a more passive role. For example, Austin et al.~\cite{austin2016smarthome} propose an in-situ monitoring system to assess well-being within one's home. Their intent is to help mitigate social biases and other inaccuracies that may impact measures made within traditional healthcare settings. To do this, their system leverages cameras, microphones, and other sensors stationed around the house so that data can be collected organically.

\textbf{Privacy Concerns:} To provide adequate health services, data from remote healthcare technology may be retained long-term so that users' habits, trends, and changes in normal health statuses can be identified. Data collected via remote healthcare technology may be stored in several places including directly on device's hardware, a separate device's hardware, in cloud-based servers, or a mix of all places to formulate a dataset that will inform healthcare treatment~\cite{siwicki2020remotedata}. It is feasible data collected by remote healthcare technology may be added to a patient's wider electronic health records (EHRs) once shared with a healthcare provider. This can be concerning as hospital data breaches in Canada~\cite{LaGrassa_2023}, and other parts of the world~\cite{vimalachandran2018preserving}, have emphasized persistent challenges in preserving patient data privacy~\cite{nowrozy2024ehr}. 

In Canada, the Privacy Act~\cite{privacyact2019} and the Personal Information Protection and Electronic Documents Act (PIPEDA)~\cite{pipeda2019} outline how the government and private-sector organizations can collect, use, and disclose personal information such as health data and medical history. As of now, remote healthcare devices may collect personal data outside of current legislation~\cite{privacycommissioner2014wearable}; how to govern this technology and adequately protect patients' data is still being investigated~\cite{cihi2023virtual}.

Literature exploring user privacy perspectives on remote healthcare technology identify a number of key themes including a desire for discretion, protection of health data, and control over how data is collected and used~\cite{wilkowska2011perception,berridge2022control,frik2020model}. These studies suggest that values relating to these themes vary depending on identity characteristics. In this paper, we closely consider age and healthcare scenario as contextual factors influencing users' needs in remote healthcare technology. 

Related studies explore age differences in users' privacy perspectives outside of healthcare. For example, when considering the privacy-protecting strategies of older adults (65+), Huang et al.'s~\cite{huang2018surfing} suggest that participants may avoid using online services if they have unresolved privacy concerns. Further, Anaraky et al.~\cite{anaraky2021disclose} explored differences in privacy related decision-making processes between older (65+) and younger adults (18-34). Their findings suggested that older adults were goal-driven and made privacy decisions based on estimated long-term outcomes. Conversely, younger adults more frequently relied on ``model-free'' ways of thinking that were adjustable to situations and environments.

When considering remote healthcare specifically, Wilkowska et al.'s~\cite{wilkowska2011perception} findings suggest differences in users' priorities; older participants were more likely to be concerned with maintaining their health rather than data security. Lorenzen-Huber et al.'s~\cite{lorenzen2011privacy} explored factors impacting perceptions of technology designed to support the health and well-being of people 65 years old and older. Participants found the technology supportive of their personal autonomy and particularly useful in emergency situations. Further, participants desired active, granular control over how their data would be shared in various contexts.

Frik et al.~\cite{frik2019privacy,frik2020model} similarly focus on the highly context-dependent decisions individuals aged 65+ make when deciding how to approach healthcare technology. Participants were not usually willing to share sensitive data collected by technology like healthcare monitoring systems. However, they were more likely to want to share this information if they perceived the data receiver as ``benevolent.'' 

\section{Methodology}
Our study was approved by our university's Research Ethics Board. We recruited adults who were comfortable completing a survey in English and who were eligible to access healthcare services in Canada. The study followed a between-subject design: each participant was assigned to one of four healthcare scenarios. We illustrate a respondent's experience completing our survey in Figure~\ref{fig:surveyflow}. Our survey is available in the appendix. It consisted of 50 questions across the four blocks summarized below. We administered two versions of our survey; one survey was hosted online via Qualtrics and the other was completed on paper. To ensure an adequate representation of the older adult population, we partnered with community groups to disseminate our paper surveys; therefore, paper survey participants were mainly adults over 50. 

To avoid overburdening participants, we implemented minor differences in the healthcare scenario section of the paper survey; both versions asked the same questions except when asking participants with whom they would share data collected by remote healthcare technology. For the online survey, we repeat this question once per data type within the scenario block. For the paper survey, we ask this question once at the end of the scenario block. We used larger font sizes and spacing between questions to ensure easy legibility. When the data sharing question was repeated, the paper survey was over 30 pages long. Thus, we modified the sharing question to reduce the survey to a reasonable length. Further, the online survey included programmed display logic whereas the paper survey simulated this logic through instructions for participants to skip a question depending on their answers. On both versions, participants could select ``Prefer not to answer'' for any given question. 

\subsection{Survey}
Participants were presented with the same questions in the first, third, and fourth block. In the second block, questions varied depending on the assigned healthcare scenario. We summarize the four healthcare scenarios in Table~\ref{table:scenarios}. Our methods in distributing participants between scenarios depended on the survey version. For online respondents, we used the Qualtrics randomizer to assign equal numbers of participants to the four health scenarios. For paper respondents, we organized survey packages to evenly distribute the scenarios once handed out. On both versions, we asked demographic questions at the end of the survey including basic information such as age, gender, race, and participants' living arrangements. 

While writing the survey, we used Microsoft Word's reading level calculator to ensure the survey content was at an appropriate reading-level for a general audience~\cite{paz2009readability}. This process included adjusting sentence structure and minimizing medical and technical terminology that may be distracting or confusing. To confirm the survey was easy to understand and seemed manageable, we piloted the survey with 3 working-age adults and 2 older adults before launch.

We briefly described remote healthcare technology as, ``...technology you can use at home to add to your typical healthcare services.'' We also briefly described their assigned healthcare scenario. For example, the chronic condition scenario description was: ``Imagine your doctor said that you have diabetes, and you will have it for the rest of your life. Alongside typical services, remote healthcare technology could help manage your chronic condition like diabetes.'' 

\begin{figure*}[tb]
    \centering
    \includegraphics[scale=0.3]{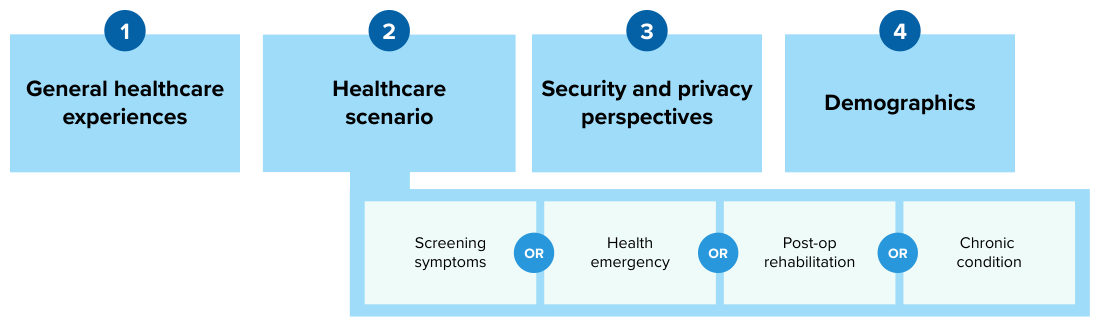}
    \caption[Survey flow]{A participant's flow through the four survey blocks.  Each participant was assigned one of four remote healthcare scenarios.}
    \label{fig:surveyflow}
\end{figure*}

\textbf{General Healthcare Experiences Block:} First, we define remote healthcare technology as \emph{technology that can be used at home in addition to typical healthcare services} in the survey. This helps account for a variety of potential solutions and to avoid bias towards a specific solution like a wearable or a monitoring system. Next, we ask participants about their general healthcare experiences and the modes (e.g., in-person, telephone, virtual video consult) in which they access healthcare services. We asked participants separate questions for experiences relating to before COVID-19 and those since the outbreak of the pandemic so that they could elaborate on any differences.

\textbf{Healthcare Scenarios Block:} Related literature~\cite{lorenzen2011privacy,czaja2019designing,frik2019privacy,frik2020model} highlights the context-driven nature of decision making when using healthcare technology, so we formulated hypothetical scenarios for participants to consider while answering the questions. While brainstorming survey scenarios, we considered a variety of conditions that varied in health severity and could impact users' potential usage of remote healthcare technology. For example, the ``Screening symptoms'' scenario represents situations where participants might experience mild symptoms and use remote healthcare for a short time to determine whether they should seek further medical assistance. In contrast, the ``Chronic condition'' scenario describes a situation wherein a participant may face relatively more critical, long-term health issues and rely on remote healthcare technology for more complex use-cases. For each scenario, we used common ailments, like a stroke or diabetes, as examples health conditions. These scenarios served as opportunity for participants to consider the use of technology within different contexts and circumstances; we do not aim to make claims about managing specific health conditions but rather broadly explore different contexts. 
\begingroup
    \setlength{\tabcolsep}{3pt}
\begin{table}[tb]
\centering
\footnotesize
\begin{tabular}{p{1.5cm} p{4.5cm} p{5.5cm}}
\hline
\textbf{Scenario} & \textbf{Potential Design Intention} & \textbf{Scenario Description in Survey} \\
\toprule
Screening symptoms & User is experiencing mild symptoms of a viral illness. Remote healthcare technology is available to identify the illness and whether they require further medical attention. & \emph{Imagine you have a fever, cough, and some sinus congestion. Alongside typical services, remote healthcare technology could help you figure out an illness and whether you need to see a doctor.} \\
\midrule
Health \newline emergency & ``Always on'' remote healthcare technology is available to detect emergencies when they happen and support the user through health emergencies. & \emph{Imagine you are at home, and you think you are having a stroke. Alongside typical services, remote healthcare technology could detect emergencies and get you help.} \\
\midrule
Post-op \newline rehab & User has recovered from recent heart surgery. Remote healthcare technology is available to support through cardiac rehabilitation activities. & \emph{Imagine that you have recovered from heart surgery. Your doctor recommends rehabilitation to strengthen your heart. Alongside typical services, remote healthcare technology could help you through rehabilitation activities.}\\
\midrule
Chronic \newline condition & User has been diagnosed with early stages of a chronic health condition. Remote healthcare technology is available to monitor symptoms and manage health. & \emph{Imagine your doctor said that you have diabetes, and you will have it for the rest of your life. Alongside typical services, remote healthcare technology could help manage your chronic condition like diabetes.} \\
\bottomrule
\end{tabular}
\caption[Summary of survey scenarios]{Participants were assigned to one scenario. Each scenario had a specific technology design goal. We include the blurb provided on the survey.}
\label{table:scenarios}
\end{table}
\endgroup

For each scenario, we presented the same questions to participants. First, on a 5-point Likert scale ranging from very unlikely to very likely, we asked participants to describe how likely they were to use remote healthcare technology to supplement their usual healthcare services with in the context of the scenario presented. Then, on a 5-point Likert scale ranging from very uncomfortable to very comfortable, we asked participants to describe their level of comfort with said remote healthcare technology collecting various types of data.

We formulated the following five data types based on those commonly collected by remote healthcare technology~\cite{siwicki2020remotedata}: (1) \textbf{Identifiable video data} such as facial recognition or video recordings; (2) \textbf{Anonymous video data} such as gesture recognition or stick-figure animations for movement tracking; (3) \textbf{Audio data} that can be used for voice-controlled assistants or ambient sound monitoring; (4)\textbf{Vital signs data} such as heart rate, blood pressure, and other information about basic bodily function; (5)\textbf{Activities and wellness data} such as daily activity, weight, posture and movement analysis.

We defined each data type and gave an example of the kinds of data that might be collected for the assigned scenario. Both paper and online participants were asked about their comfort with having remote healthcare technology collect each data type. If participants responded that they were uncomfortable or very uncomfortable with a specific type of data, they were then asked to identify their concerns from a list we formulated based on concerns frequently mentioned in related literature~\cite{frik2020model,Ray2021,berridge2022control}. We included space to describe any concerns we had not listed. 

Lengthy questionnaires can have lower completion rates and poorer quality responses than shorter questionnaires~\cite{galseic2009questionnaire}. To mitigate potential respondent fatigue, we introduced branching in our survey. For example, we focused on the concerns of individuals who identified discomfort. If participants responded that they were neutral, comfortable, or very comfortable with the data type being collected for the given scenario, we asked them to select with whom they would share this information from a list of entities including their healthcare providers, friends, family, and device manufacturer. Participants could select ``No one'' if they would not share the data collected by the device. We asked online survey participants about their sharing preferences for each data type separately, and we asked paper survey participants about their sharing preferences in general.

At the end of the scenario block, we provided an open-text box to elaborate further on any thoughts or concerns relating to using remote healthcare technology in the provided scenario. 

\textbf{Security and Privacy Block:} We asked questions relating to participants' general digital security attitudes and behaviours. We used a modified version of the SA-6~\cite{faklaris2019sa6} --  a validated tool comprised of 6 Likert scale questions associated with security-related behaviours and intentions. We adjusted statements to suit a lower reading level, and excluded one question relating to having diligent security protection practices. During validation completed by Faklaris et al.~\cite{faklaris2019sa6}, the SA-6's internal consistency remained strong (alpha = 0.81) when this item was deleted.  

We also presented the following three privacy statements to assess participants' general privacy perspectives: \emph{``I am concerned about threats to my privacy online''} (Priv1); \emph{``I already take steps to protect my privacy''} (Priv2); and, \emph{``I would give up some privacy to use a service''} (Priv3). We pulled from related literature for these statements. For example, \emph{Priv1} is from the Global Information Privacy Concern (GIPC) scale~\cite{naresh2004iuipc}. \emph{Priv2} was generated by Colnago et al.~\cite{colnago2022concern} for their study on users' interpretation of privacy statements. Using Colnago et al.'s~\cite{colnago2022concern} recommendations for effective privacy statements, we formulated \emph{Priv3} to explore privacy trade-offs.

\subsection{Recruitment}
We used multiple recruitment strategies. Some recruitment initiatives included adults of all ages while others (e.g., paper survey) targeted adults over 50 to ensure sufficient representation of this population in our sample. Compensation varied based on the recruitment method. We detail our various recruitment methods below.

We recruited eligible adult participants via Prolific in waves to monitor the distribution of participants. Data was collected through Qualtrics, and we leveraged Prolific's pre-screening feature to target participants of specific demographics (i.e., ages, genders) each round to balance our sample. Participants from Prolific were compensated at Prolific's recommended rate of £3.00GBP (\$5.00CAD) for a survey of this length. Participants recruited via Prolific tend to have experience participating in online research projects, technological literacy, reliable internet access, and tend to be younger. We thus supplemented our sample by recruiting in two other ways.

We recruited via social media (e.g., Facebook, Twitter), our personal networks, and local community outreach. Individuals recruited via social media completed the Qualtrics survey, and could enter a raffle for a \$50 CAD gift card (odds of winning were 1:50). Contact information for the raffle was collected separately to maintain anonymity. 

We also reached out to older adult groups in-person by visiting public libraries and community centres with specialized programming for adults over 50. Upon invitation, we gave a brief presentation to community groups about our survey goals. We delivered manila envelopes containing blank paper surveys and additional leaflets to enter the raffle. Participants were given as much time as they needed to complete the survey. We added these paper responses to our digital database before shredding them. 

\subsection{Dataset and Statistical Analysis Methods}
We received a total of 728 responses to our survey. We eliminated digital responses wherein the metadata looked suspicious. For example, we removed multiple entries that had the exact same responses and duplicate start/end timestamps. We hypothesize that these suspicious entries were attempts to win the survey raffle as they came in response to a social media recruitment initiative. After data cleaning, we had 384 valid responses; 63\% of the final set of responses were from participants recruited via Prolific, 32\% from participants recruited via other online recruitment efforts, and 5\% from participants recruited via community outreach. We analyzed 384 valid responses in the final dataset. 

We used the following statistical analyses to address our research questions. To address RQ1, we use descriptive statistics to summarize responses in related survey blocks exploring participants' comfort with data collection (Section~\ref{comfort}), related concerns (Section~\ref{concerns}), and sharing preferences (Section~\ref{sharing}). To address RQ2, we conducted inferential statistics on our dataset to compare for effects of two independent variables: (1) participant age group, and (2) assigned healthcare scenario. The Kruskal-Wallis test is a non-parametric alternative to a one-way ANOVA that can be used to test more than two samples. We use this test to compare differences between our ordinal data (i.e., Likert-scale responses) and totals of categorical data (e.g., concerns, sharing recipients). In instances where statistically significant results were found, we followed up with pairwise Wilcoxon rank sum test (i.e., Mann-Whitney's U test) with Bonferroni correction to compare our independent samples and identify where the differences occurred. We consider tests with $p<0.05$ to be statistically significant.

\section{Results}
\label{results}
We organize results by survey block. Within each, we identify how results relate to our research questions (labelled RQ1 and RQ2). We primarily focus on statistically significant results on the inferential statistics exploring age group and scenario effects relating to RQ2. Non-significant findings are included in Table~\ref{tab:allresults-age} and Table~\ref{tab:allresults-scenario} in the appendix. We interpret the overall survey results in Section~\ref{sec:discussion}.

\subsection{Participants}
\label{Participants}
\begin{table}[tb]
    \centering
    \footnotesize
    \begin{tabular}{lrrrrr}
    \toprule
    \textbf{Group} & \textbf{Total} & \textbf{Women} & \textbf{Men} & \textbf{Non-Binary} & \textbf{Undisclosed} \\
    \midrule
   18-34 & 71 & 34 & 33 & 3 & 1 \\
   35-49 & 76 & 37 & 39 & 0 & 0\\
   50-64 & 144 & 78 & 62 & 2 & 2\\
   65+ & 93 & 58 & 33 & 2 & 0 \\
    \midrule
    Symptoms  &  96 & 43 & 48 & 2 & 0   \\
    Rehab     & 97 & 54 & 42 & 2 & 0 \\
    Emergency & 98 & 45 & 51 & 1 & 0 \\
    Chronic   & 93 & 64 & 27 & 2 & 3 \\
    \midrule
    \emph{Across all} & \emph{384} & \emph{206} & \emph{168} & \emph{7} & \emph{3} \\
    \bottomrule
    \end{tabular}
    \caption[Ages and genders of participants]{Summary of gender distribution between age groups and scenarios.}
    \label{tab:age-gender}
\end{table}

We provide a high-level summary of our participants in Table~\ref{tab:age-gender}. Participant ages ranged from 20 to 93 with a median age of 53. Fifty-four per cent  of all participants identified as women. Further, 75\% of participants were white, 15\% were Asian, the remaining 10\% of participants were Black, Indigenous, or mixed race. In terms of education, 22\% of participants had at most a high school diploma, 49\% of the participants' highest degree was a college or university certificate, and 27\% had a graduate or professional degree. 90\% of participants lived in a private residence like a house, condo, or apartment. 82\% of participants lived with other people including a partner or other relatives like parents, siblings, or children. 

\subsection{Healthcare Experiences}
Participants briefly described their current healthcare experiences. We summarize the commonalities in responses. Many participants reported mixed experiences receiving healthcare in Canada, e.g., \emph{``I have received some excellent care and also some appallingly bad care [...]''}. When describing good experiences, participants said their needs were met and that working with healthcare providers was generally positive. Further, many participants appreciated the \textbf{coverage of costs}, \emph{``I feel that we are tremendously blessed with our healthcare system. I feel fortunate that we can see the doctor, access blood tests, get X-rays... all covered by our healthcare system.''} Many participants also appreciated the \textbf{speed of first response teams}, \emph{``I find that emergency services do respond quite quickly [...].''} 

However, \textbf{long wait times} was the predominant concern, e.g., \emph{``The wait to see a doctor is very long and wait times for diagnostic results and procedures are getting long.''} and \emph{``Accessing specialist care is very frustrating as it takes far too long.''} Participants also frequently mentioned having \textbf{too limited time with care providers}, e.g., \emph{``Family doctors are too rushed to deal with my questions. If you have more than a couple of questions you're out of luck.''} Participants mentioned that they \textbf{more frequently experienced poor care} due to the strains put on the healthcare system from COVID-19. Many mentioned that care has degraded in quality, e.g., \emph{``There has been a major crisis going on in the health care system since COVID-19 pandemic.''}.

Few participants reflected on remote healthcare technology in this section. Some described \textbf{previous experience with remote healthcare services}, e.g., \emph{``I have received healthcare services via in-person, telephone consultations and through video chat sessions.''} While others considered their \textbf{general impressions} of receiving healthcare remotely, e.g., \emph{``I am open to using it if it were appropriate; I have phoned my doctor a few times in the past with small questions so I imagine it would be similar to that.''} and \emph{``I think I prefer a face-to-face consultation for most situations.''}

\subsection{Likelihood of Use} Using a 5-point Likert scale, participants ranked their likelihood to use remote healthcare technology to capture a first impression immediately following the definition of remote healthcare technology and description of the scenario. 

{\textbf{\emph{RQ1.}} Providing further context for our first research question, the overall mean participant likelihood of use rating was 3.82 out of 5 ($SD=1.21, MD=4$)\footnote{Note: We acknowledge diverging views on how Likert scale data should be reported. We report medians because it is ordinal data, but also include means and standard deviations (though this ordinal data may not be normally distributed).} suggesting that participants were generally slightly positive in adopting remote healthcare technology. 

\emph{\textbf{RQ2.} Age Groups:} Our tests revealed a statistically significant effect of age group on likelihood of use($\chi^2(3) = 16.5, p < 0.01$).  Post-hoc analysis found significant differences (i) between the \emph{35-49} and \emph{65+} age groups (p < 0.01, r = 0.19) and (ii) between the \emph{50-64} and \emph{65+} age groups (p = 0.01, r = 16). Both the \emph{35-49} age group ($M = 4.1, SD = 1.04, MD = 4.0$) and the \emph{50-64} age group ($M = 3.9, SD = 1.04, MD = 4.0$)  had a higher means than the \emph{65+} age group ($M = 3.3, SD = 1.46, MD = 4.0$). 

\emph{\textbf{RQ2.} Healthcare Scenarios:} The mean ratings were highest among participants in the \emph{chronic condition} scenario ($M=4.0, SD = 1.06, MD = 4.0$) and lowest for the \emph{post-op rehabilitation} scenario ($M=3.7, SD = 1.40, MD = 4.0$). However, our tests revealed no statistically significant effect of scenario on participants' likelihood of use ratings. 

\subsection{Comfort with Data Collection}
\label{comfort}
In the Healthcare Scenario block, we introduced five data types. Using 5-point Likert scales, participants ranked their comfort with having remote healthcare technology collect each data type for their given scenario. 

\textbf{\emph{RQ1.}} The majority of participants were neutral or positive about the collection of each data type. Participants seemed most comfortable with the collection of \emph{wellness and activities} data ($M = 3.8, SD = 1.15, MD = 4.0$) and least comfortable with \emph{identifiable video} data ($M = 3.2, SD = 1.25, MD = 4.0$).

\emph{\textbf{RQ2.} Age Groups:} We found differences in responses per age group for only one data type: \emph{identifiable video} data ($\chi^2(3) = 9.19, p = 0.03$). Our post-hoc test showed significant differences between the \emph{18-34} and \emph{50-64} age groups (p = 0.04, r = 0.13). Participants in the \emph{18-34} age group ($M=2.9, SD = 1.22, MD = 3.0$) were less comfortable with the collection of \emph{identifiable video} data than those in the \emph{50-64} age group ($M=3.4, SD = 1.11, MD = 4.0$). 

\emph{\textbf{RQ2.} Healthcare Scenarios:} We found a statistically significant effect of scenario on comfort ratings for the collection of only one data type: \emph{audio} data ($\chi^2(3) = 12.8, p < 0.01$). Post-hoc analysis showed significant differences between the \emph{chronic condition} scenario and the \emph{emergency} scenario (p < 0.01, r = 0.82) for \emph{audio} data. Participants were less comfortable with the collection of \emph{audio} data for the \emph{chronic condition} scenario ($M=3.2, SD=1.20, MD=4.0$) than for the \emph{emergency} scenario ($M=3.8, SD=1.08, MD=4.0$). No other pairwise comparisons were statistically significant.

\subsection{Concerns with Data Collection}
\label{concerns}
Participants who said they were \textbf{uncomfortable} or \textbf{very uncomfortable} with the collection of a given data type were then asked about their concerns relating to the data type. We further analyse the subset of participants ($N=192$) who were asked about concerns below.

To begin, we provided four concerns and an ``other'' option; participants could select all that applied. Only 9\% of participants selected the ``other'' option. We reviewed these responses but they did not detail new examples of explicit risks. Instead, these participants expressed generalized privacy concerns or discomfort with the specific data type being collected. For example, \emph{``I think it would be weird if something was always listening to me,''} and \emph{``It feels too much like being tracked and that makes me uncomfortable.''} Others voiced general concern about whether this data could lead to an accurate diagnosis, e.g., \emph{``A stethoscope is [a] more useful tool to diagnose lung disease. Listening to a cough is not enough.''}

\begin{table}[tb]
\centering
\footnotesize
\begin{tabular}{l p{2cm} p{2cm} p{2cm} p{2cm} p{2cm} p{2cm}}
\toprule
\textbf{Group} &
  \textbf{Identifiable video} &
  \textbf{Anonymous video} &
  \textbf{Audio} &
  \textbf{Vital signs} &
  \textbf{Wellness \& activity} \\
\midrule
18-34     & \cellcolor[HTML]{FFFC9E}\textbf{41}\% & 23\% & 25\% & 7\%  & 10\%  \\
35-49     & \cellcolor[HTML]{FFFC9E}\textbf{30}\% & 25\% & 25\% & 29\% & 21\% \\
50-64     & \cellcolor[HTML]{FFFC9E}\textbf{24}\% & 13\% & \cellcolor[HTML]{FFFC9E}\textbf{24}\% & 18\% & 13\% \\
65+       & \cellcolor[HTML]{FFFC9E}\textbf{40}\% & 29\% & 24\% & 19\% & 16\% \\
\midrule
Symptoms  & \cellcolor[HTML]{FFFC9E}\textbf{37}\% & 25\% & 23\% & 18\% & 20\%  \\
Rehab     & \cellcolor[HTML]{FFFC9E}\textbf{36}\% & 17\% & 23\% & 17\% & 11\%  \\
Emergency & \cellcolor[HTML]{FFFC9E}\textbf{28}\% & 20\% & 15\% & 20\% & 13\%  \\
Chronic   & 29\% & 23\% & \cellcolor[HTML]{FFFC9E}\textbf{35}\% & 19\% & 14\%  \\
\midrule
\emph{Across all} &
  \cellcolor[HTML]{FFFC9E}\emph{\textbf{32\%}} &
  \emph{21\%} &
  \emph{24\%} &
  \emph{18\%} &
  \emph{15\%} \\
\bottomrule
\end{tabular}
\caption{[Concerns] Percentage of participants expressing any concern with each data type. We highlight the most concerning data types per group.}
\label{tab:concern-percent-bydatatype}
\end{table}

\definecolor{Gray}{gray}{0.94} %create custom colour
\begin{table}[tb]
\footnotesize
\centering
\begin{tabular}{l p{1.5cm} p{1.5cm} p{1.5cm} p{1.5cm} p{1.5cm}}
\toprule
\textbf{Group} &
  \textbf{Don't Know} &
  \textbf{Don't Want} &
  \textbf{Misuse} &
  \textbf{Storage} &
  \textbf{Other} \\ 
\midrule
18-34     & 24\% & 32\% & \cellcolor[HTML]{FFFC9E}\textbf{41}\% & 34\% & 8\%   \\
35-49     & 22\% & 21\% & \cellcolor[HTML]{FFFC9E}\textbf{37}\% & 36\% & 1\%   \\
50-64     & 13\% & 16\% & \cellcolor[HTML]{FFFC9E}\textbf{30}\% & 24\% & 8\% \\
65+       & 23\% & 16\% & \cellcolor[HTML]{FFFC9E}\textbf{33}\% & 29\% & 15\%  \\
\midrule
Symptoms  & 19\% & 17\% & \cellcolor[HTML]{FFFC9E}\textbf{34}\% & 33\% & 8\%  \\
Rehab     & 22\% & 28\% & \cellcolor[HTML]{FFFC9E}\textbf{35}\% & 28\% & 9\%  \\
Emergency & 18\% & 19\% & \cellcolor[HTML]{FFFC9E}\textbf{33}\% & 27\% & 5\%  \\
Chronic   & 17\% & 17\% & \cellcolor[HTML]{FFFC9E}\textbf{34}\% & 29\% & 11\%  \\
\midrule
\emph{Across all} &
  \emph{19\%} &
  \emph{20\%} &
 \cellcolor[HTML]{FFFC9E} \emph{\textbf{34}\%} &
  \emph{29\%} &
  \emph{8\%} \\ 
\bottomrule
\end{tabular}
\caption{[Concerns] Percentage of participants selecting each concern (for at least one data type). We highlight the most popular concern per group.}
\label{tab:concern-percent-byconcerntype}
\end{table}

\begin{table}[tb]
\centering
\footnotesize
\begin{tabular}{l p{1.5cm} p{1.3cm} p{1.3cm} p{1.5cm} p{1.5cm}|p{1.6cm}}
\toprule
\textbf{Data Type} & \textbf{Don't Know} & \textbf{Don't Want}  & \textbf{Misuse} & \textbf{Storage} & \textbf{Other} & \textbf{Total Concerns}\\\midrule
Identifiable video    & 11\% & 14\% & \cellcolor[HTML]{FFFC9E}\textbf{35}\% & 32\% & 9\%  & 219 \\
Anonymous video       & 9\%  & 21\% & \cellcolor[HTML]{FFFC9E}\textbf{29}\% & 24\% & 12\% & 141 \\
Audio                 & 19\% & 18\% & \cellcolor[HTML]{FFFC9E}\textbf{34}\% & 25\% & 7\%  & 163 \\
Vital signs           & 14\% & 17\% & \cellcolor[HTML]{FFFC9E}\textbf{33}\% & 30\% & 3\%  & 138 \\
Wellness \& activity & 9\%  & 18\% & \cellcolor[HTML]{FFFC9E}\textbf{33}\% & 32\% & 3\%  & 91  \\
\midrule
\emph{Across all} & \emph{14\%}      & \emph{17\%}      & \cellcolor[HTML]{FFFC9E}\emph{\textbf{33}\%}   & \emph{29\%}    & \emph{7\%}   & \emph{752}   \\ \bottomrule
\end{tabular}
\caption{[Concerns] Percentage of participants expressing each concern per data type. We highlight in yellow the concern identified by the most participants per data type.  The ``Total Concerns'' column sums all concerns per data type.}
\label{tab:concerntype-bydatatype}
\end{table}

\emph{\textbf{RQ1.}} We summarize the percentage of participants who expressed concern per data type in Table~\ref{tab:concern-percent-bydatatype}, and per type of concern in Table~\ref{tab:concern-percent-byconcerntype}.  In general, participants had the most concerns with remote healthcare technology collecting \emph{identifiable video} data (32\%) and they were most concerned with data being \emph{misused} (34\%). 

We cross-tabulate concerns with data types in Table~\ref{tab:concerntype-bydatatype}.  Across all data types, participants were most concerned with \emph{misuse} of data. For example, $35\%$ of all 219 concerns for \emph{identifiable video} related to data \emph{misuse}. We then calculated 9 totals for each participant. First, we counted the total number of concerns identified per data type (i.e., giving 5 totals per participant). For example, a participant who expressed concerns about data \emph{misuse} and data storage for \emph{audio} data would have a total count of 2 for \emph{audio} data. Next, we counted the total number of times each concern was selected across all data types (i.e., giving 4 more totals per participant, leaving out `Other'). For example, a participant who was concerned about data \emph{misuse} for \emph{audio}, \emph{vital signs}, and \emph{identifiable video} data would have a count of 3 for data \emph{misuse}. To measure the effect of age group and scenario, we conducted Kruskal-Wallis tests on these totals. 

\emph{\textbf{RQ2. } Age Groups:} We explored effects of age group on total concerns per data type. Our tests revealed statistically significant differences for only one data type: \emph{identifiable video} data ($\chi^2(4) = 11.76, p = 0.02$). Post-hoc tests revealed statistically significant differences between the \emph{18-34} age group and the \emph{50-64} age group (p = 0.03, r = 0.14). 
The \emph{18-34} age group ($M=1.9, SD=0.88, MD=2.0$) expressed more concerns relating to \emph{identifiable video} data than the \emph{50-64} age group ($M=1.6, SD=0.85, MD=1.0$). Additionally, we explored the effect of age group on total number of times each concern was selected across all data types. Our tests revealed no statistically significant differences.

\emph{\textbf{RQ2. } Healthcare Scenarios:} 
Our tests revealed no effect of scenario on the total number of concerns identified per data type, nor on the total number of times each concern was selected across all data types. 

\begin{figure}[tb]
     \centering     
     \begin{subfigure}{\textwidth}
        \centering
        \includegraphics[scale=0.6]{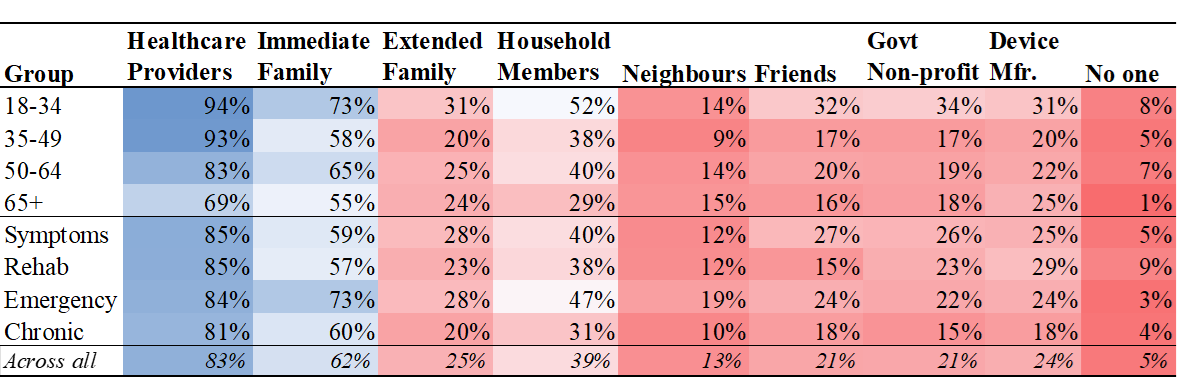}
        \caption{Across all data types, the percentages of participants who would share per recipient. }
        \label{fig:share-alltypes}
     \end{subfigure}

     \begin{subfigure}{\textwidth}
         \centering
         \includegraphics[scale=0.6]{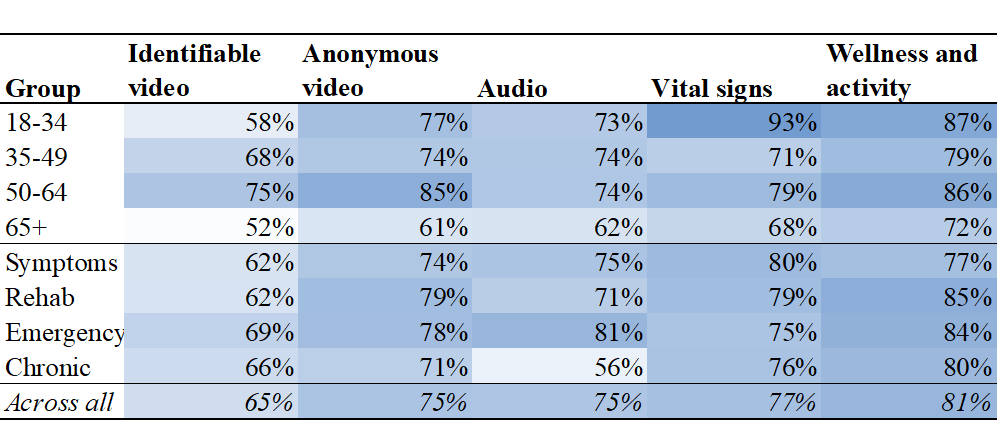}
         \caption{The percentages of participants who would share a particular data type. 
         %Cells are coloured on a scale from darkest red (0\%) to darkest blue (100\%).
         }
         \label{fig:share-eachtype}
     \end{subfigure}
    \caption[Sharing preferences]{[Sharing preferences] Higher percentages (blue) indicate more participants would share data; red indicates fewer participants. Cells are coloured from darkest red (0\%) to darkest blue (100\%).}
\end{figure}

\subsection{Sharing Preferences}
\label{sharing}

Participants who were either \textbf{neutral}, \textbf{comfortable} or \textbf{very comfortable} with the collection of a data type were asked with whom they would share data. Participants could select as many recipients as applicable. For the online survey, we asked this question per data type.\footnote{To reduce the number of questions, we did not ask about about sharing for each data type on the paper version of the survey.} In total, 363 online participants were asked about data sharing; analysis in this section is from this subset of participants. 

\textbf{\emph{RQ1.}} Figure~\ref{fig:share-alltypes} summarizes which recipients participants would share collected data. In general, participants most often selected to share information gathered by remote healthcare technology with healthcare providers, followed by immediate family. Only a few participants were comfortable with data collection but unwilling to share the data being collected with anyone (i.e., selected ``no one'' (N=29)). Table~\ref{fig:share-eachtype} summarizes sharing preferences per data type. Although more than half of participants agreed to share data with at least one recipient, participants were most inclined to share \emph{wellness and activities} data and \emph{vital signs} data, and least inclined to share \emph{identifiable video} data.

\emph{\textbf{RQ2. } Age Groups:} We considered differences in the total number of recipients per data type. Our tests revealed revealed an effect of age group on the number of recipients for two data types: \emph{identifiable video} data ($\chi^2(3) = 10.11, p = 0.02$) and \emph{anonymous video} data ($\chi^2(3) = 12.09, p < 0.01$). Post-hoc tests for \emph{identifiable video} data showed significant differences between the \emph{18-34} and \emph{50-64} age groups (p = 0.03, r = 0.14), and between the \emph{18-34} and \emph{65+} age groups (p = 0.04, r = 0.14). On average, the \emph{18-34} age group selected to share \emph{identifiable video} data with more recipients ($M = 2.9, SD=1.52, MD = 2.0$) than the \emph{50-64} age group ($M = 2.2, SD=1.11, MD = 2.0$) and the \emph{65+} age group ($M = 2.2, SD = 1.42, MD = 2.0$). Post-hoc tests for \emph{anonymous video} data showed significant differences between the \emph{18-34} and \emph{50-64} groups ($p < 0.01, r = 0.17$), where those in the \emph{18-34} group would share anonymous video data with more recipients ($M = 3.2, SD = 1.94, MD = 3.0$) than the \emph{50-64} group ($M = 2.3, SD = 1.39, MD = 2.0$). 

\emph{\textbf{RQ2. } Healthcare Scenarios:} In Figure~\ref{fig:share-alltypes}, we also describe the average number of recipients per scenario. For example, participants in the \emph{chronic} scenario selected fewer recipients across all data types ($M = 8.1, SD = 4.78, MD = 7.0$) and participants in the \emph{emergency} scenario reported a willingness to share with the most recipients ($M = 10.9, SD = 7.69, MD = 10.0$). However, our tests revealed no significant effect of scenario on the number of recipients per data type. 

\subsection{Security and Privacy Attitudes}
We used a modified version of SA-6~\cite{faklaris2019sa6} for our study. To avoid priming participants on other parts of the survey, we asked these questions about security and privacy at the end. We determined each participant's SA-6 score by converting their Likert scale responses to numbers (i.e., 1= strongly disagree, 5 = strongly agree) and then calculating a geometric mean so that we could account for respondents who responded ``Prefer not to answer.'' The highest possible SA-6 score is 5 which suggests a high level of security awareness~\cite{faklaris2019sa6}. 
Participants' mean SA-6 score was 3.54 ($SD = 0.91, MD = 3.5$) suggesting they were somewhat security aware. Further, participants generally agreed with two privacy statements: ``I am concerned about threats to my privacy online'' ($M = 4.2, SD = 0.99, MD = 4.0$) and ``I already take steps to protect my privacy'' ($M = 4.2, SD = 0.95, MD = 4.0$). Participants were more neutral towards ``I would give up some privacy to use a service'' ($M = 3.0, SD = 1.17, MD = 3.0$). More details on SA-6 responses are including in Table~\ref{tab:sa6priv}.

\emph{Age Groups:} We found an effect of age group on mean SA-6 scores ($\chi^2(3) = 14.08, p < 0.01$). Post-hoc tests showed statistically significant differences between the \emph{18-34} and \emph{50-64} age groups ($p < 0.01, r = 0.19$), where the \emph{50-64} age group ($M=3.8, SD=0.77, MD=4.0$) was more security aware than the \emph{18-34} age group ($M=3.4, SD=0.81, MD=3.2$). We also found an effect of age group on agreement with the statement, ``I am concerned about threats to my privacy online'' ($\chi^2(3) = 10.33, p = 0.02$). Post-hoc tests showed significant differences between the \emph{18-34} and \emph{65+} age groups (p = 0.02, r = 0.15) where the \emph{65+} age group ($M=4.2, SD=0.83, MD=4.0$) indicated more concern than the \emph{18-34} age group ($M=3.8, SD=1.25, MD=4.0$). We found no significant effect of age group for the other privacy statements. 

\emph{Healthcare Scenarios:} We found an effect of scenario on mean SA-6 scores ($\chi^2(3) = 30.68, p < 0.01$). Post-hoc tests showed statistically significant differences between (i) the \emph{emergency} and \emph{chronic} scenarios ($p < 0.01, r = 0.23$) and between (ii) the \emph{emergency} and \emph{symptoms screening} scenarios ($p < 0.01, r = 0.25$). Participants in the \emph{emergency} scenario were less security aware ($M=3.1, SD=0.82, MD=3.0$) than those in both the \emph{chronic} ($M=3.7, SD=0.87, MD=4.0$) and \emph{symptoms screening} ($M=3.7, SD=0.89, MD=4.0$) scenarios. We found no effect of scenario on any of the privacy statements.

\section{Discussion}
\label{sec:discussion}
Our survey findings provide us with insight into participants' comfort with data collection, sharing preferences, and potential privacy concerns related to remote healthcare technology. We discuss our key takeaways and provide considerations for the design of future remote healthcare technology for people in Canada.

\textbf{\emph{RQ1: What are users' perspectives on data collected and shared by remote healthcare technology?}}
Earlier literature~\cite{frik2019privacy,ray2022olderadults} considers general data collection through remote healthcare technology; we extend the literature by considering nuances in privacy concerns for each data type.

\textbf{Identifiable video data differs from other types of data.} Overall, the majority of survey participants were neutral or positive about the collection of each data type. This suggests that users may find it generally acceptable for remote healthcare technology to collect data to support their existing healthcare services. However, our findings suggest that users may accept the collection of some data types more than others. In particular, participants were least comfortable with the collection of identifiable video data and this trend held across the different scenarios, including health emergencies. Thus, it appears that identifiable video data is particularly sensitive for remote healthcare purposes.

Discomfort with video data collection aligns with related literature ~\cite{zheng2018smarthome,bugeja2016smarthome,austin2016smarthome} exploring users' concerns with video recording capabilities in smart home devices such as video doorbells and smart baby monitors. Within the context of remote healthcare technology, the collection of video data can help serve critical health needs such as facilitating emergency care by detecting falls or seizures. Despite potential life-saving benefits, our findings suggest that privacy cynicism~\cite{hoffmann2016privacy} relating to particular data types towards video data persists.

We theorize that participants' persistent aversion to identifiable video in this context may stem from reasons not fully explored in smart home technology research. For example, remote healthcare technology users may not be willing to disclose video data because the collection of identifiable video data is not as common as other types of data that are typically collected via popular health tracking devices (e.g., a personal EKG monitor or a FitBit) and applications (e.g., a glucose monitoring application or the iPhone health app); users may not be accustomed to sharing identifiable video data to manage their health and well-being. Per Contextual Integrity~\cite{nissenbaum2004contextual}, such violations of norms and societal expectations influence individual privacy expectations and may ultimately discourage the adoption of technology. 

Further, users in Canada may be particularly sensitive to collecting video data because general awareness of risks relating to video surveillance~\cite{cbc2024yorkpeel,cbc2023winnipeg} and deepfakes~\cite{dubinski2024kids,karadeglija2024hate} has increased. This increased sensitivity may impact users' privacy calculus~\cite{culnan1999information}. The potential costs of disclosing video data with remote healthcare technology may outweigh perceived health benefits; i.e., individuals may view risks of disclosing video data with remote healthcare technology to be greater than health risks that could be mitigated by collecting this data. 

\textbf{Potential benefits of sharing.} When participants were comfortable with the collection of a particular data type, they were also likely willing to share the data with at least one recipient. As an abstraction, imagine placing recipients in a series of concentric circles. We would have healthcare providers in the innermost circle, followed by immediate family, then household members. The outermost circle would contain neighbours, friends, device manufacturers, government, and non-profit organizations. Participants were most likely to share with recipients in the innermost circle and least likely to share with those in the outermost circle.

Related literature~\cite{frik2020model} suggests that a remote healthcare technology user is most likely to share data with individuals who can most benefit the user. Our findings suggest that participants may be inclined to share data with recipients who have direct impact to their general health and well-being. If future remote healthcare technology involves sharing data with recipients outside of inner circles, users may be more willing to share data if they are provided with a clear explanation of how sharing may improve their healthcare services.

\textbf{\emph{RQ2: How do perspectives and considerations vary depending on contexts such as a user's age and healthcare scenario?}} Participants were mostly in favour of adopting remote healthcare technology. We found a few differences between age groups and scenarios.

\textbf{Older adults and adoption.} In Canada, 85\% of older adults prefer to live and care for their well-being in their homes and communities~\cite{nrc2020challenge}. Adoption of remote healthcare technology could thus be critical for these older adults (i.e., allowing them to age in place). However, participants aged 65+ were significantly less inclined to use remote healthcare technology than those of younger age groups. This aligns with related literature also showing low adoption rates for older adults~\cite{beer2011acceptance,cimperman2016acceptance}. 

Future remote healthcare technology will need to address the issues impeding adoption for people in this age group.
Encouragingly, participants within the \emph{65+} age group did not have significantly more concerns than the other age groups, suggesting that improved adoption is feasible. Further, the trends in likelihood of use ratings for those approaching their senior years (i.e., participants in the \emph{50-64} age group) suggest that adoption rates among seniors may soon increase.

Designers of remote healthcare technology should carefully consider \emph{generational factors} that can impact adoption rates. For example, people aged 60+ have historically been the least likely to use the internet; this phenomenon, coined as the ``grey digital divide''~\cite{hunsaker2018internetuse}, has been observed around the world~\cite{berkowsky2018challenges,hou2022internetsocialparticipation}. However, as of 2020, internet use among adults aged 65 and above in Canada had increased to 77\%, suggesting that the gap is closing~\cite{wavrock2022covid}.

\textbf{Healthcare scenarios.} 
We extend previous literature by exploring how user privacy preferences concerns may vary depending on specific healthcare scenarios. We anticipated that the four healthcare scenarios would represent different contexts, each having their own impact on participants' perspectives. 

We found a few differences between scenarios and those were aligned with expectations. For example, participants were more comfortable sharing audio data in the emergency scenario than in the chronic scenario. Considering privacy calculus~\cite{culnan1999information}, individuals may  weigh the potential benefit of sharing audio data during an urgent life-threatening situation to be worth the privacy disclosure, but not when managing a situation with less imminent risks. 

We speculate on the potential reasons for uncovering fewer differences between scenarios than expected. First, participants may consider `healthcare' to be the primary context and may not discriminate between individual scenarios. Alternatively, our scenario descriptions and questions may not have been sufficient to capture the nuances. Future work, such as user interviews, on this topic can continue to explore potential effects of contextual factors including those not covered by our survey like social norms and cultural influences that can impact users' needs and expectations in this technology.

\subsection{Limitations and Future Work}
We used branching in our survey to mitigate respondent fatigue; however, this may impact the generalizability  of our findings. For example, we did not ask sharing questions to participants who were uncomfortable with data collection and thus our sharing insights do not include those perspectives. We provided open-ended questions for participants to elaborate on their perspectives, yet a survey format limits further probes about the rationale behind responses. 

We recruited participants via Prolific and community outreach. This sampling technique was helpful for collecting initial insights. However, this technique may have introduced sampling bias and selection bias. Further, we provided examples of data that could be collected to help participants contextualize the role of remote healthcare technology within a healthcare scenario. This necessarily introduced some variability between data type descriptions and may have impacted the effect of scenario on our data.

Future work could include studies exploring (i) how other characteristics (e.g., gender, ethnicity) impact perspectives on remote healthcare technology, (ii) perspectives on relevant data protection policies and whether these policies impact adoption, and (iii) preferences relating to sharing recipients more closely. Our ongoing work includes an interview study with older adults to further explore this demographic's needs and concerns with remote healthcare technology.

\section{Conclusion}
Remote healthcare technology has the potential for positive impact on healthcare systems; however, implementations must consider how to handle sensitive data in ways which meet users' privacy expectations. We explored participants' perspectives on remote healthcare technology and nuances specific to age groups and healthcare scenarios. Our findings suggest that the majority of participants were likely to adopt remote healthcare technology and that they were receptive to allowing each of the five types of data to be collected for healthcare purposes. We found nuanced differences in responses based on participants' age, and few differences based on assigned healthcare scenario. For example, despite shrinking generational gaps relating to technology use, participants 65 years and older were significantly less inclined to adopt remote healthcare technology. Additionally, most participants were willing to disclose sensitive data when faced with imminent health risks. Finally, we contextualized these findings by discussing how they compare to user preferences with other technology and theorized why common aversions to collecting certain data remain despite the potential health benefits of using remote healthcare technology. 

\section*{Acknowledgements}
The authors acknowledge financial support for this research from the Human-Centric Cybersecurity Partnership, which is a SSHRC Partnership Grant.

%
% ---- Bibliography ----
%
% BibTeX users should specify bibliography style 'splncs04'.
% References will then be sorted and formatted in the correct style.
%
\bibliographystyle{splncs04}
\bibliography{main.bib}
%
%%
%% If your work has an appendix, this is the place to put it.
\appendix

\begin{table}
\centering
\footnotesize
\begin{tabular}{>{\raggedright\arraybackslash}p{3.8cm}l|>{\raggedleft\arraybackslash}p{1.3cm}   >{\raggedleft\arraybackslash}p{1.3cm}   >{\raggedleft\arraybackslash}p{1.3cm}   >{\raggedleft\arraybackslash}p{1.3cm}   >{\raggedleft\arraybackslash}p{1.3cm}   >{\raggedleft\arraybackslash}p{1.3cm}   }
\toprule
 &
   &
  \multicolumn{6}{c}{Wilcoxon Rank Sum} \\
\textbf{} &
  KW &
  18-34 vs 35-49 &
  18-34 vs 50-64 &
  18-34 vs 65+ &
  35-49 vs 50-64 &
  35-49 vs 65+ &
  50-64 vs 65+ \\ 
  %\hline
  \midrule
\emph{\quad Likelihood to use remote healthcare technology} &
  \cellcolor[HTML]{FFFC9E}\textgreater{}0.01 &
  0.40 &
  1.00 &
  0.27 &
  0.90 &
  \cellcolor[HTML]{FFFC9E}\textgreater{}0.01 &
  \cellcolor[HTML]{FFFC9E}0.01 \\ 
  %\hline
  \midrule
\emph{Comfort ratings with collection of...} &
   &
   &
   &
   &
   &
   &
   \\
\emph{\quad Identifiable video data} &
  \cellcolor[HTML]{FFFC9E}0.02 &
  0.16 &
  \cellcolor[HTML]{FFFC9E}0.04 &
  1.00 &
  1.00 &
  1.00 &
  1.00 \\
\emph{\quad Anonymous video data} &
  0.24 &
  - &
  - &
  - &
  - &
  - &
  - \\
\emph{\quad Audio data} &
  0.98 &
  - &
  - &
  - &
  - &
  - &
  - \\
\emph{\quad Vital signs} &
  0.75 &
  - &
  - &
  - &
  - &
  - &
  - \\
\emph{\quad Wellness and activity} &
  0.22 &
  - &
  - &
  - &
  - &
  - &
  - \\ 
\midrule
\emph{Total concerns with collection of...} &
   &
   &
   &
   &
   &
   &
   \\
\emph{\quad Identifiable video data} &
  \cellcolor[HTML]{FFFC9E}0.02 &
  1.00 &
  \cellcolor[HTML]{FFFC9E}0.03 &
  1.00 &
  1.00 &
  1.00 &
  0.07 \\
\emph{\quad Anonymous video data} &
  0.26 &
  - &
  - &
  - &
  - &
  - &
  - \\
\emph{\quad Audio data} &
  0.75 &
  - &
  - &
  - &
  - &
  - &
  - \\
\emph{\quad Vital signs} &
  0.83 &
  - &
  - &
  - &
  - &
  - &
  - \\
\emph{\quad Wellness and activity} &
  0.35 &
  - &
  - &
  - &
  - &
  - &
  - \\ 
\midrule
\emph{``I do not know why this information would need to be collected.''} &
  0.07 &
  - &
  - &
  - &
  - &
  - &
  - \\
\emph{``I do not want people to know this information.''} &
  0.10 &
  - &
  - &
  - &
  - &
  - &
  - \\
\emph{``I worry this information may be misused.''} &
  0.62 &
  - &
  - &
  - &
  - &
  - &
  - \\
\emph{``I worry this information may not be properly protected when stored.''} &
  0.41 &
  - &
  - &
  - &
  - &
  - &
  - \\ 
\midrule
\emph{Total participants who would share with...} &
   &
   &
   &
   &
   &
   &
   \\
\emph{\quad Healthcare Provider} &
  0.17 &
  - &
  - &
  - &
  - &
  - &
  - \\
\emph{\quad Immediate family} &
  0.15 &
  - &
  - &
  - &
  - &
  - &
  - \\
\emph{\quad Extended family} &
  0.53 &
  - &
  - &
  - &
  - &
  - &
  - \\
\emph{\quad Household members} &
  \cellcolor[HTML]{FFFC9E}0.02 &
  0.23 &
  0.73 &
  \cellcolor[HTML]{FFFC9E}0.02 &
  1.00 &
  1.00 &
  0.50 \\
\emph{\quad Neighbours} &
  0.70 &
  - &
  - &
  - &
  - &
  - &
  - \\
\emph{\quad Friends} &
  0.05 &
  - &
  - &
  - &
  - &
  - &
  - \\
\emph{\quad Government/Non-profit} &
  0.05 &
  - &
  - &
  - &
  - &
  - &
  - \\
\emph{\quad Device manufacturer} &
  0.27 &
  - &
  - &
  - &
  - &
  - &
  - \\
\emph{\quad No one} &
  0.15 &
  - &
  - &
  - &
  - &
  - &
  - \\ 
\midrule
\emph{Total participants who would share...} &
   &
   &
   &
   &
   &
   &
   \\
\emph{\quad Identifiable video data} &
  \cellcolor[HTML]{FFFC9E}0.02 &
  0.06 &
  \cellcolor[HTML]{FFFC9E}0.03 &
  \cellcolor[HTML]{FFFC9E}0.04 &
  1.00 &
  1.00 &
  1.00 \\
\emph{\quad Anonymous video data} &
  \cellcolor[HTML]{FFFC9E}\textgreater{}0.01 &
  0.08 &
  \cellcolor[HTML]{FFFC9E}\textgreater{}0.01 &
  0.28 &
  1.00 &
  1.00 &
  1.00 \\
\emph{\quad Audio data} &
  0.09 &
  - &
  - &
  - &
  - &
  - &
  - \\
\emph{\quad Vital signs} &
  0.05 &
  - &
  - &
  - &
  - &
  - &
  - \\
\emph{\quad Wellness and activity} &
  0.11 &
  - &
  - &
  - &
  - &
  - &
  - \\ 
\bottomrule
\end{tabular}
\caption{A summary of the statistical tests we ran to explore the effects of \textbf{age group} on survey responses. We list p-values for our Kruskal-Wallis (KW) tests; when significant, we ran pairwise comparisons using Wilcoxon rank sum (i.e., Mann Whitney's U) tests. All significant results (>0.05) are highlighted in yellow and discussed further in Section \ref{results}.}
\label{tab:allresults-age}
\end{table}

%%%%%%%%%%%%%%%%%%%%%%%%%%%%%%%%%%%%%%%%%%%%%%%%%%%%%%%%%%%%
\begin{table*}
\centering
\footnotesize
\begin{tabular}{>{\raggedright\arraybackslash}p{3.8cm}l|>{\raggedleft\arraybackslash}p{1.3cm}   >{\raggedleft\arraybackslash}p{1.3cm}   >{\raggedleft\arraybackslash}p{1.3cm}   >{\raggedleft\arraybackslash}p{1.3cm}   >{\raggedleft\arraybackslash}p{1.3cm}   >{\raggedleft\arraybackslash}p{1.3cm}   }
\toprule
&      & \multicolumn{6}{c}{Wilcoxon Rank Sum} \\
\textbf{} &  KW &  18-34 vs 35-49 &  18-34 vs 50-64 &  18-34 vs 65+ &  35-49 vs 50-64 &  35-49 vs 65+ &  50-64 vs 65+ \\ 
%\hline
\midrule
\emph{Likelihood to use remote healthcare technology} & 0.41 & -    & -    & -    & -    & -   & -   \\ 
%\hline
\midrule
\emph{Comfort ratings with collection of...}          &      &      &      &      &      &     &     \\
\emph{\quad Identifiable video data} &  \cellcolor[HTML]{FFFC9E}0.02 & 0.16 &  \cellcolor[HTML]{FFFC9E}0.04 &  1.00 &  1.00 &  0.46 &  0.19 \\
\emph{\quad Anonymous video data}                       & 0.24 & -    & -    & -    & -    & -   & -   \\
\emph{\quad Audio data}                                     & 0.98 & -    & -    & -    & -    & -   & -   \\
\emph{\quad Vital signs}                                    & 0.22 & -    & -    & -    & -    & -   & -   \\
\emph{\quad Wellness and activity}                          & 0.75 & -    & -    & -    & -    & -   & -   \\ 
%\hline
\midrule
\emph{Total concerns with collection of...}           &      &      &      &      &      &     &     \\
\emph{\quad Identifiable video data}                        & 0.78 & -    & -    & -    & -    & -   & -   \\
\emph{\quad Anonymous video data}                           & 0.73 & -    & -    & -    & -    & -   & -   \\
\emph{\quad Audio data}                                     & 0.26 & -    & -    & -    & -    & -   & -   \\
\emph{\quad Vital signs}                                    & 0.75 & -    & -    & -    & -    & -   & -   \\
\emph{\quad Wellness and activity}                          & 0.36 & -    & -    & -    & -    & -   & -   \\ 
%\hline
\midrule
\emph{``I do not know why this information would need to be collected.''} &
  0.23 &  - &  - &  - &  - &  - &  - \\
\emph{``I do not want people to know this information.''} &
  0.75 &  - &  - &  - &  - &  - &  - \\
\emph{``I worry this information may be misused.''}   & 0.74 & -    & -    & -    & -    & -   & -   \\
\emph{``I worry this information may not be properly protected when stored.''} &
  0.68 &  - &  - &  - &  - &  - &  - \\ 
%\hline
\midrule
\emph{Total participants who would share with...}     &      &      &      &      &      &     &     \\
\emph{\quad Healthcare Provider}                            & 0.17 & -    & -    & -    & -    & -   & -   \\
\emph{\quad Immediate family}                               & 0.15 & -    & -    & -    & -    & -   & -   \\
\emph{\quad Extended family}                                & 0.34 & -    & -    & -    & -    & -   & -   \\
\emph{\quad Household members}                              & 0.13 & -    & -    & -    & -    & -   & -   \\
\emph{\quad Neighbours}                                     & 0.36 & -    & -    & -    & -    & -   & -   \\
\emph{\quad Friends}                                        & 0.08 & -    & -    & -    & -    & -   & -   \\
\emph{\quad Government/Non-profit}                          & 0.08 & -    & -    & -    & -    & -   & -   \\
\emph{\quad Device manufacturer}                            & 0.25 & -    & -    & -    & -    & -   & -   \\
\emph{\quad No one}                                         & 0.25 & -    & -    & -    & -    & -   & -   \\ 
%\hline
\midrule
\emph{Total participants who would share...}          &      &      &      &      &      &     &     \\
\emph{\quad Identifiable video data}                        & 0.98 & -    & -    & -    & -    & -   & -   \\
\emph{\quad Anonymous video data}                           & 0.11 & -    & -    & -    & -    & -   & -   \\
\emph{\quad Audio data}                                     & 0.13 & -    & -    & -    & -    & -   & -   \\
\emph{\quad Vital signs}                                    & 0.26 & -    & -    & -    & -    & -   & -   \\
\emph{\quad Wellness and activity}                          & 0.22 & -    & -    & -    & -    & -   & -   \\ 
%\hline
\bottomrule
\end{tabular}
\caption{A summary of the statistical tests we ran to explore the effects of \textbf{healthcare scenario} on survey responses. We list p-values for our Kruskal-Wallis (KW) tests; when significant, we ran pairwise comparisons using Wilcoxon rank sum (i.e., Mann Whitney's U) tests. All significant results (>0.05) are highlighted in yellow and discussed further in Section \ref{results}.}
\label{tab:allresults-scenario}
\end{table*}

%%% CUSTOM TABLE %%%
\definecolor{Gray}{gray}{0.94} %create custom colour
\newcolumntype{z}{>{\columncolor{Gray}}l} %z for grey cols, left-aligned
\newcolumntype{y}{>{\columncolor{Gray}}r} %y for grey cols, right-aligned
\newcolumntype{x}{>{\columncolor{Gray}}c} %y for grey cols, center
\begin{table}[tb]
\centering
\footnotesize
\begin{tabular}{lyyyrrryyyrrr}
\toprule
 & 	\multicolumn{3}{x}{\textbf{SA-6 Score}} & \multicolumn{3}{c}{\textbf{Priv1}} & \multicolumn{3}{x}{\textbf{Priv2}} & \multicolumn{3}{c}{\textbf{Priv3}} \\ 
\textbf{} & \textbf{Mean} & \textbf{SD} & \textbf{MD} & \textbf{Mean} & \textbf{SD} & \textbf{MD} & \textbf{Mean} & \textbf{SD} & \textbf{MD} & \textbf{Mean} & \textbf{SD} & \textbf{MD} \\ 
\midrule
18-34 & 3.3 & 0.85 & 3.3 
& 3.8 & 1.25 & 4.0 
& 4.1 & 1.02 & 4.0 
& 3.2 & 1.20 & 3.0 \\
35-49 & 3.6 & 1.00 & 3.6 
& 4.3 & 0.93 & 4.0 
& 4.4 & 0.72 & 5.0 
& 3.2 & 1.05 & 3.0 \\
50-64 & 3.7 & 0.86 & 4.0 
& 4.2 & 0.83 & 4.0 
& 4.2 & 0.96 & 4.0 
& 3.0 & 1.14 & 3.0 \\
65+ & 3.5 & 0.90 & 3.5 
& 4.3 & 1.00 & 5.0 
& 4.1 & 1.04 & 4.0 
& 2.8 & 1.28 & 3.0 \\ 
\midrule
Symptoms & 3.7 & 0.89 & 4.0 
& 4.3 & 0.83 & 4.0 
& 4.1 & 1.04 & 4.0 
& 3.1 & 1.11 & 3.0 \\ 
Rehab & 3.6 & 0.93 & 4.0
& 4.2 & 1.04 & 5.0 
& 4.2 & 0.81 & 4.0 
& 2.9 & 1.18 & 3.0 \\
Emergency & 3.2 & 0.82 & 3.0 
& 4.2 & 0.95 & 4.0 
& 4.3 & 1.02 & 4.0 
& 3.2 & 1.17 & 3.0 \\
Chronic & 3.7 & 0.87 & 4.0 
& 4.0 & 1.12 & 4.0 
& 4.2 & 0.95 & 4.0 
& 2.9 & 1.24 & 3.0 \\
\midrule
\emph{Across all} 
& \emph{3.5} & \emph{0.91} & \emph{3.5} 
& \emph{4.2} & \emph{0.95} & \emph{4.0}
& \emph{4.2} & \emph{0.95} & \emph{4.0}
& \emph{3.0} & \emph{1.17} & \emph{3.0} \\ 
\bottomrule
\end{tabular}
\caption[Participant results for the modified SA-6 and privacy statements]{[Security and privacy] Descriptive statistics for the modified SA-6 scores and 5-point Likert responses to 3 privacy statements.}
\label{tab:sa6priv}
\end{table}
%%% CUSTOM TABLE %%%

\clearpage

       

\section*{Supplementary material}

**Remote Healthcare – Online Survey – Chronic Condition example**

Remote healthcare technology is technology you can use at home to add to your typical healthcare services. We will ask about your preferences for using this technology. No experience with remote healthcare technology is needed. Your answers are anonymous. You can skip questions you don't want to answer. Your answers will be saved, so you can finish the survey later if you need a break. Do not share names or personal details.

<u>Your Healthcare Experiences</u>

Think about healthcare such as medical care, pharmacy, physical therapy, psychotherapy.

1. What is your general experience receiving healthcare in Canada? (Suggested: 1-2 sentences or "Prefer not to answer.")
2. Since COVID-19, how has your experience with healthcare changed? (Options: Now is worse than before COVID-19, About the same, Now is better than before COVID-19, Prefer not to answer)
   a) What has made your healthcare experience worse? (Suggested: 1-2 sentences or "Prefer not to answer.")
   b) What has made your healthcare experience better? (Suggested: 1-2 sentences or "Prefer not to answer.")
3. Before the COVID-19 pandemic, how often did you use these methods for getting healthcare? (Options: Never, Sometimes, Most of the time, Always, I don't know, Prefer not to answer)
   a) In-person
   b) Telephone
   c) Virtual video consult
   d) Email
   e) Instant message / chat (for example, Facebook messenger, Whatsapp, Text message/SMS)

4. Since the COVID-19 outbreak, how often have you used these methods for getting healthcare? (Options: same as above)

<u>When answering the next questions, consider this scenario:</u>

Imagine your doctor said that you have diabetes, and you will have it for the rest of your life. Alongside typical services, remote healthcare technology could help manage your chronic condition like diabetes.

1. Have you experienced a chronic condition such as diabetes or otherwise? (Options: Yes, No, I don't know, Prefer not to answer)
2. Would you use remote healthcare technology to help manage a chronic condition? (Options: Very unlikely, Somewhat unlikely, Neither likely nor unlikely, Somewhat likely, Very likely, Prefer not to answer)

<u>Data Collection</u>

Remote healthcare technology can collect different information.

Identifiable video data is video information where a computer or human can recognize you. For example, it could check your skin for infections.

1. To help manage a chronic condition, I would be _____ if remote healthcare technology collected identifiable video data. (Options: Very uncomfortable, Somewhat uncomfortable, Neither comfortable nor uncomfortable, Somewhat comfortable, Very comfortable)
   a. Why would you be uncomfortable if remote healthcare technology collected identifiable video data to help manage a chronic condition? Select all that apply. (Options: I do not know why this information would be needed for this scenario; I do not want people to know this information; I

worry this information may be misused; I worry this information may not be protected when stored, Other concern(s), Prefer not to answer)

    b. Tell us about your other concerns with identifiable video data being collected to help manage a chronic condition.  [Open-text box]

2. To help manage a chronic condition, I would share identifiable video data with _____. Select all that apply. If you would not share, select "No one." (Options: Healthcare Providers, Immediate family, Extended family, Household members, Neighbours, Friends, Government / Non-profit, Device manufacturer, No one, Prefer not to Answer)

*This block is repeated once per following data type:

- Anonymized video data is video information where a computer or human cannot recognize who is in the video. For example, it could see you as a stick-figure to check if you are having trouble walking.
- Audio data is information detected by a microphone. For example, it could hear if you fall or faint.
- Activities and wellness data is general information that describes you. For example, it could help you track your sugar intake.
- Vital signs data is information that describes your basic bodily functions. For example, it could monitor your heart if your blood sugar gets low.

In this section, please tell us more about your digital security and privacy perspectives.

(Options: Strongly disagree, Somewhat disagree, Neither agree nor disagree, Somewhat agree, Strongly agree, Prefer not to answer)

1. I look for opportunities to learn about security measures that are relevant to me.
2. I am very motivated to take all the steps needed to keep my online data and accounts safe.
3. I am often interested in articles about security threats.
4. I always pay attention to experts' advice about the steps I need to take to keep my online data and accounts safe.
5. I am very knowledgeable about all the steps needed to keep my online data and accounts safe.
6. I am concerned about threats to my privacy online.
7. I already take steps to protect my privacy.
8. I would give up some privacy to use a service.

Tell us more about yourself.

This information will help us understand who took our survey. Reminder: Your answers are anonymous. You can skip questions you don't want to answer. Your answers will be saved, so you can finish the survey later if you need a break. Do not share names or personal details.

1. What year were you born? [Year of Birth dropdown]
2. Gender: I am [Options: A woman, A man, Non-binary, Not listed (Please specify), Prefer not to say]
3. What is your race and/or ethnicity? [Open text box]
4. Where in Canada do you live? [Options: Alberta, British Columbia, Manitoba, New Brunswick, Newfoundland and Labrador, Northwest Territories, Nova Scotia, Nunavut, Ontario, Prince Edward Island, Quebec, Saskatchewan, Yukon, Somewhere outside of Canada, Prefer not to answer]
5. What level of schooling have you completed? Select all that apply. [Options: High school diploma or equivalency certificate, College or university degree, Graduate or professional degree, None of the above, Prefer not to answer]
6. What language are you most comfortable speaking? [Options: English, French, Not listed (Please specify), Prefer not to answer]
7. How do you usually access the Internet? Select all that apply. [Options: Home computer (For example: desktop, laptop), Tablet (For example: iPad), Smartphone, Not listed (Please specify), Prefer not to answer]

8. Where do you live or spend most of your time? [Options: In a private residence (For example: house, condo, apartment), In an independent living residential accommodation (For example: on campus, retirement home), In an assisted living facility (For example: nursing home), Not listed (Please specify), Prefer not to answer]

9. Who lives with you? Select all that apply. [Options: No one I live alone, A partner, Other relatives (For example: parents, siblings, children), Roommates, Other people (Please specify general relationship), Prefer not to answer]

10. Who helps you with your health and wellness needs (For example: cleaning the house, hygiene)? Select all that apply. [Options: No one, I take care of myself; Friend or family, roommate who lives with me; Friend or family who lives outside my home; A paid caregiver or service (For example: nurse, attendant, meal service; A volunteer caregiver or service (For example: community member); Not listed (Please specify general relationship); Prefer not to answer]

11. Has your race, ethnicity, or spoken language impacted your experiences receiving healthcare services in Canada? [Options: Yes, it has positively impacted my experience; Yes, it has negatively impacted my experience; No, it has had no impact; Prefer not to answer]

    a. Optional: Please share more on how your race, ethnicity, or spoken language may have impacted your experiences receiving healthcare services. Do not share specific names, places, or other identifying information.

12. Optional: If you have comments, concerns, or suggestions relating to this survey, please leave them below.


\end{document}